\documentclass[floatfix,aps,prl,showpacs,amsmath,nofootinbib,twocolumn]{revtex4}
\usepackage{graphicx}
\newcommand{\be}{\begin{equation}}
\newcommand{\ee}{\end{equation}}

\begin{document}

\title{Can superhorizon perturbations drive the acceleration of the Universe?}
\author{\'Eanna \'E. Flanagan}\email{eef3@cornell.edu}
\affiliation{
Laboratory for Elementary Particle Physics, Cornell University, Ithaca, NY 14853, USA.}

\date{\today}

\begin{abstract}

It has recently been suggested that the acceleration of the Universe
can be explained as the backreaction effect of superhorizon
perturbations using second order perturbation theory.
If this mechanism is correct, it should also apply to a hypothetical,
gedanken universe in which the subhorizon perturbations are absent.
In such a gedanken universe it is possible to compute the deceleration
parameter $q_0$ measured by comoving observers using local covariant
Taylor expansions rather than using second order perturbation theory.
The result indicates that second order corrections to $q_0$ are
present, but shows that if $q_0$ is negative then its magnitude is
constrained to be less than or of the order of the square of the
peculiar velocity on Hubble scales today.  We argue that since this
quantity is constrained by observations to be small compared to unity,
superhorizon perturbations cannot be responsible for the acceleration
of the Universe.

\end{abstract}


\maketitle

The observed acceleration of the expansion of the Universe
\cite{SN,wmap} is a profound mystery.  It is usually explained by
positing a new form of matter
with negative pressure, so-called dark energy \cite{Peebles}, or by a
modification of general relativity at large distance scales
\cite{modifiedgravity}.  Recently it has been suggested that the
acceleration is instead driven by the backreaction effect of
inflation-generated superhorizon perturbations via second order
perturbation theory \cite{Kolb,Barausse}.
The backreaction of perturbations had earlier been
studied in other contexts by Brandenberger and collaborators
\cite{Brandenberger} and others \cite{Nambu}.

The basic idea is the following.  As is well known, scalar perturbation modes
whose wavelengths today are smaller than $\lambda \sim 8 \, {\rm
Mpc}$ have entered the nonlinear regime and are responsible for
galaxies and galaxy clusters, while longer wavelength modes and in particular
superhorizon modes ($\lambda \agt 3 \, {\rm Gpc}$) are still in the
linear regime today.  That is, the fractional
density perturbation due
to these modes is small compared to unity, so each individual mode
evolves with high accuracy according to the linearized equations.
Nevertheless the net effect of the backreaction from all the
superhorizon modes can still be significant.  If we denote by
$\varepsilon = \delta \rho/\rho \sim 10^{-4}$ the fractional density
perturbation on Hubble scales today, then one naively expects second order
corrections to be of order $\varepsilon^2 \sim 10^{-8}$ which is
negligible.  However, a more refined estimate for some particular
second order effects obtained from integrating over all
the superhorizon modes \cite{Barausse} gives the scaling
$\varepsilon^2 F(k_{\rm min}, k_{\rm max})$, where $k_{\rm min}$
and $k_{\rm max}$ are the minimum and maximum comoving wavenumbers of
the scalar perturbation spectrum integrated over.  Here $F$ is a
function for which $F \to \infty$
as $k_{\rm min} \to 0$.  Thus, if the spectrum extends to sufficiently
large wavelengths (as would be generated by a sufficiently long period
of inflation), second order effects can become significant.
In particular, Refs.\ \cite{Kolb,Barausse} argue that such effects can
give rise to an effective cosmological constant and drive the
present acceleration of the Universe.

A potential difficulty with this idea is that it appears to be
in conflict with
the locality and causality of general relativity.
Specifically, consider the finite spacelike hypersurface ${\cal
V}$ given by the interior of the intersection of our past
lightcone with the spacelike hypersurface of some fixed redshift $z=z_0$.
We will take $z_0=5$ say so that the observed supernova used as
evidence of the universe's acceleration are to the future of ${\cal
  V}$.  Then, the initial data within ${\cal V}$ are sufficient to determine all
the observations we make; in particular the observable effect of
superhorizon perturbations must be encoded in this initial
data\footnote{Note that this initial data does not itself contain
  superhorizon modes: ${\cal V}$ is a finite spherical
region whose comoving radius is $\sim 1.8 H_0^{-1}$ for $z_0=5$.  So
for modes which are superhorizon today, ${\cal V}$ extends across only
a small fraction of the mode wavelength at $z=z_0$.}.  In addition
observations strongly constrain this initial data: we know that the
geometry of ${\cal V}$ on large scales can be accurately modeled as a
Friedman-Robertson-Walker (FRW)
background plus fractional perturbations of order $\varepsilon\sim
10^{-4}$.  Thus the issue is whether or not the space of initial data for the
gravitational and matter fields on ${\cal V}$ contains enough freedom,
given the observational constraints, to mimic the effects of dark
energy when no such dark energy is present in the matter stress energy
tensor.

In this paper we shall argue that the freedom is insufficient.
More precisely, we focus on the deceleration parameter $q_0 = -
a(t) {\ddot a}(t) / {\dot a}(t)^2$, where $a(t)$ is the scale factor,
whose measured value is $q_0 \sim -0.5$.  We will argue that (i) The
local spatial curvature within ${\cal V}$ is unconstrained and can be
altered by second order effects, giving rise to changes in
$q_0$ that could in principle be of order unity.  However this effect
cannot change a positive value of $q_0$ to a
negative value of $q_0$.  (ii) Non-isotropy and non-homogeneity of the
initial data in ${\cal V}$ on large scales can give rise to negative
values of $q_0$, as suggested by Refs.\ \cite{Moffat}.  However the
magnitude of this effect is
constrained to be of order the square of the velocity
perturbation on Hubble scales, which is constrained by observations of
low order multipoles of the cosmic microwave background to
be small compared to unity.

\noindent
{\it Method of analysis:}
If the superhorizon perturbation mechanism for driving
acceleration is correct, it should apply not just to our Universe but
also to other hypothetical universes.  For ease
of analysis, we will analyze a fictitious, gedanken universe which
differs from ours only in that the perturbation spectrum at
early times is modified to suppress the perturbations which are
subhorizon today.  In this universe the subhorizon perturbations are
negligible today, while the superhorizon perturbations
are taken to be the same as those used in Refs.\ \cite{Kolb,Barausse}.
A demonstration that the superhorizon
perturbation mechanism does not work in this context is fairly strong
evidence that it cannot work in our Universe.  The only possible loophole is
the possibility that subhorizon perturbations somehow play an
important role, which does not seem to be indicated by the analyses of
Refs.\ \cite{Kolb,Barausse} \footnote{In addition momentum conservation $k
\approx k_1 + k_2$ rules out second order corrections to low spatial
frequency observables ($k \alt H_0$) from interactions between very
subhorizon modes $k_1 \gg H_0$ and superhorizon modes $k_2 \alt H_0$.}.

In this gedanken universe, the length and time scales over which the
gravitational and matter fields are varying are all of order
$H_0^{-1}$ or larger.  This allows us to perform an analysis in a
local region using Taylor series expansions of the Einstein and
hydrodynamic equations, which is much simpler than second order
perturbation theory about a FRW background.

We model the matter source by the fluid stress energy tensor
$T_{\alpha\beta} = (\rho + p) u_\alpha u_\beta +
p g_{\alpha\beta}$, where $\rho$, $p$, $u^\alpha$ and
$g_{\alpha\beta}$ are the density, pressure, 4-velocity and metric.
Consider a comoving observer at an event ${\cal P}$.  Such an observer
can measure the redshift $z$ and luminosity distance ${\cal L}$ of
nearby events, and thus measure the redshift-luminosity distance
relation
$
{\cal L} = {\cal L}(z,\theta,\varphi).
$
Here $\theta$ and $\varphi$ are spherical polar coordinates in the
observer's local Lorentz frame.  The dependence on these angles arises
since we are allowing general local solutions of the Einstein
equations; there is no requirement of isotropy.  For small $z$ this
relation can be expanded as \cite{Barausse}
\be
{\cal L} = A(\theta,\varphi) z + B(\theta,\varphi) z^2 + O(z^3).
\label{eq:expansion0}
\ee
We define the Hubble constant $H_0$
and deceleration parameter $q_0$ as measured by the observer by
comparison with the conventional FRW relation ${\cal L} = H_0^{-1} z
+ H_0^{-1} (1 - q_0) z^2/2 + O(z^3)$, as in Ref.\ \cite{Barausse}:
\be
H_0 \equiv \langle A^{-1} \rangle, \ \ \ \ \
q_0 \equiv 1 - 2 H_0^{-2} \langle A^{-3} B \rangle.
\label{eq:q0def}
\ee
Here the angular brackets denote an average over the angles $\theta,\varphi$.

The particular prescription (\ref{eq:q0def}) for angular averaging
is chosen
for later convenience.  Note that there is no unique prescription; one could for
example use the definitions $H_0^{-1} = \langle A \rangle$ and $q_0 =
1 - 2 H_0 \langle B \rangle$.
Observations that measure $H_0$ and $q_0$ typically assume isotropy
and therefore effectively angle-average at some stage of the analysis,
but the precise nature of the averaging is not usually discussed.  We
will argue below however that the effect of this ambiguity on the
values of $H_0$ and $q_0$ is small.

Using local Taylor series expansions we can compute $H_0$ and $q_0$ in terms
of the the density, 4-velocity and velocity gradients of the
cosmological fluid evaluated at the observer's location ${\cal P}$.
We decompose the gradient in the usual way as
\be
\nabla_\alpha u_\beta = {1 \over 3} \theta (g_{\alpha\beta} +
u_{\alpha} u_{\beta}) + \sigma_{\alpha\beta} + \omega_{\alpha\beta}
- u_\alpha a_\beta,
\label{eq:decompos}
\ee
where $\theta$, $\sigma_{\alpha\beta}$, $\omega_{\alpha\beta}$ and
$a_\alpha$ are the expansion, shear, vorticity and 4-acceleration.
For $H_0$ we find the well-known result
\be
H_0 = {1 \over 3} \theta;
\label{eq:H0ans}
\ee
the locally measured Hubble constant is just the expansion of the
fluid.  For $q_0$ we obtain
\begin{eqnarray}
q_0  &=& {4 \pi \over 3 H_0^2} (\rho + 3 p) + {1 \over 3 H_0^2}
\bigg[ a_\alpha a^\alpha +{7 \over 5} \sigma_{\alpha\beta}
  \sigma^{\alpha\beta}
\nonumber \\ \mbox{} &&
-  \omega_{\alpha\beta} \omega^{\alpha\beta}
- 2 \nabla_\alpha a^\alpha \bigg].
\label{eq:q0result}
\end{eqnarray}
Now to a good approximation in the present epoch we
have $p=0$ (assuming cold dark matter and baryons with no dark energy), which
implies from $\nabla_\alpha T^{\alpha\beta}=0$ that $a_\alpha=0$.  This yields
\be
q_0  = {4 \pi \over 3 H_0^2} \rho  + {1 \over 3 H_0^2}
\left[ {7 \over 5} \sigma_{\alpha\beta} \sigma^{\alpha\beta} -
 \omega_{\alpha\beta} \omega^{\alpha\beta} \right].
\label{eq:ans}
\ee

\noindent
{\it Discussion:}  Consider first the first term in Eq.\
(\ref{eq:ans}).  This reduces to the conventional value $q_0 = 1/2$
for a flat, matter dominated Universe when $H_0^2 = 8 \pi \rho/3$.  However in
the present context the relation $H_0^2 = 8 \pi \rho/3$
need not be satisfied; $H_0$ is instead given by Eq.\
(\ref{eq:H0ans}).  The deviation of this first term from $1/2$ is
related to the fact that the local analysis allows
spatial curvature.  If we define an effective local $\Omega_k$ by
$\Omega_k = 1 - 8 \pi \rho / (3 H_0^2)$, then we obtain for the first
term $q_0 = (1 - \Omega_k)/2$, the conventional answer for a Universe
with matter and spatial curvature.

The key point about the first term in (\ref{eq:ans}) is that it is
positive.  Hence this term cannot drive an acceleration.

Consider next the second and third terms in Eq.\ (\ref{eq:ans}), the
squared shear and squared vorticity.  These quantities have an
unambiguous operational meaning; they can be measured by the observer in her local
Lorentz frame.
We can estimate the sizes of
these terms as
$
\sigma_{ab} \sigma^{ab}, \omega_{ab} \omega^{ab} \sim (\delta v)^2 / l^2,
$
where $\delta v$ is the typical scale of peculiar velocity (deviation from
uniform Hubble flow), and $l$ is the lengthscale over which the
velocity varies.  In the present context we have $l \agt H_0$, by our
assumption that subhorizon modes are negligible, which implies that
the contribution from the second and third terms in (\ref{eq:ans}) to
$q_0$ is of order
$
\delta q_0 \sim (\delta v)^2 \sim \varepsilon \sim 10^{-4}.
$
We conclude that it is impossible in this model to achieve the measured
value $q_0 \sim -0.5$ of the deceleration parameter.

Note that the key difference between our analysis and that of Refs.\
\cite{Kolb,Barausse} is one of interpretation.
Refs. \cite{Kolb,Barausse} predict
changes to $q_0$ that are quadratic in the the first order
perturbation variables, in agreement with our Eq.\ (\ref{eq:ans}).
The new information provided by our analysis is that the quadratic
terms are in fact locally measurable and represent degrees of
freedom of the cosmological fluid rather than of the gravitational field.
In the argument above, $\delta v$ characterizes the total deviation of
the fluid velocity from an FRW background, including both
first and second order perturbations.  The contribution of the
quadratic terms in Eq.\ (\ref{eq:ans}) to $q_0$ are constrained to be
small since observations constrain the total velocity perturbation
$\delta v$.  Thus, while an order-unity renormalization of $q_0$ from
second order effects is possible in principle, our analysis implies
that such a renormalization would also require second order contributions
to the fluid velocity that violate observational bounds.

\noindent
{\it Details of derivation:} We use the local covariant expansion
formalism based on bitensors \cite{DB60}.  We denote by $x^\alpha$ the
coordinates of the event ${\cal
  P}$ where the observer is making observations, and by $x^{\alpha'}$
the coordinates of an event ${\cal Q}$ in the observer's vicinity.  We
shall mostly be interested in the case where ${\cal Q}$ is on the past
lightcone of ${\cal P}$.  We denote by $\lambda$ an affine parameter
along the geodesic $x^\alpha = z^\alpha(\lambda)$ that joins ${\cal
  Q}$ and ${\cal P}$, chosen so
that $\lambda =0$ at $Q$ and $\lambda =1$ at ${\cal P}$.
We define Synge's world function (the squared geodesic interval) via
\be
\sigma(x,x') = {1 \over 2} \int_0^1 d\lambda \,
g_{\alpha\beta}[z^\alpha(\lambda)] {d z^\alpha \over d
  \lambda}(\lambda) {d z^\beta \over d\lambda}(\lambda).
\ee
Then $\sigma_{;\alpha}(x,x') = \nabla_\alpha \sigma(x,x')$ is the
tangent to the
geodesic at ${\cal P}$.  We define $s(x,x') = - \sigma_{;\alpha}(x,x')
u^\alpha(x)$; we will use $s$ as our expansion parameter.  We define the
vector $k^\alpha$ by
\be
\sigma_{;\alpha}(x,x') = s(x,x') k_\alpha(x);
\label{eq:kdef}
\ee
this is a future directed tangent to the geodesic which is
normalized so that $k_\alpha u^\alpha =-1$ at ${\cal P}$.
We define $g_\alpha^{\ \alpha'}(x,x')$ to be the linear operator of
parallel transport from the tangent space at ${\cal Q}$ to the tangent
space at ${\cal P}$, and we define
$
{\bar u}^\alpha(x,x') =
g^{\alpha\alpha'}
(x,x') u_{\alpha'}(x').
$
This quantity can be expanded in a
local covariant Taylor series as
\begin{eqnarray}
{\bar u}^\alpha(x,x') &=& u^\alpha(x) + u^{\alpha\beta}(x)
  \sigma_{;\alpha}(x,x') \sigma_{;\beta}(x,x')
\nonumber \\
\mbox{} &&
+ {1 \over 2}
u^{\alpha\beta\gamma}(x)
  \sigma_{;\alpha}(x,x') \sigma_{;\beta}(x,x') \sigma_{;\gamma}(x,x')
\nonumber \\
\mbox{} &&
  + O(s^4).
\label{eq:expansion}
\end{eqnarray}
Using Eq.\ (B14) of Ref.\ \cite{AFO} we can evaluate the coefficients
to give $u^{\alpha\beta}(x) = - \nabla^{(\alpha} u^{\beta)}(x)$ and
$u^{\alpha\beta\gamma}(x) = \nabla^{(\alpha} \nabla^\beta u^{\gamma)}(x)$.

The measured redshift of the event ${\cal Q}$ is given by
the ratio of the inner products ${\vec k} \cdot {\vec u}$ evaluated at
${\cal Q}$ and ${\cal P}$.  Using the definition of ${\bar u}^\alpha$
this can be written as
\be
1+z = {{\bar u}^\alpha k_\alpha \over u^\alpha k_\alpha}.
\ee
Using the expansion (\ref{eq:expansion}) and the definition
(\ref{eq:kdef}) this can be written as
\be
z = u^{\alpha\beta} k_\alpha k_\beta s - {1 \over 2}
u^{\alpha\beta\gamma} k_\alpha k_\beta k_\gamma s^2 + O(s^3).
\label{eq:redshift}
\ee

Next we evaluate the luminosity distance ${\cal L}$.  This is defined
so that $4 \pi {\cal L}^2$ is the ratio between an energy emitted per
unit time isotropically at ${\cal Q}$ and an energy per unit time per
unit proper area received at ${\cal P}$:
\be
{dE \over dt}({\cal Q}) = 4 \pi {\cal L}^2 {dE \over dt d^2A}({\cal
  P}).
\ee
These quantities can be evaluated using the
geometric optics approximation to the scalar (or
Maxwell) wave equation \cite{Barausse}.  The stress-tensor for the
radiation field is $T_{\alpha\beta} = A^2 l_\alpha l_\beta$, where
$l_\alpha$ is defined as being the set of null vectors at ${\cal Q}$
normalized according to $l_{\alpha'} u^{\alpha'} =-1$, and then
extended along ${\cal Q}'s$ future light cone using the geodesic
equation.  We define the affine parameter ${\bar \lambda}$ by ${\vec
  l} = d/d {\bar \lambda}$.
The normalization conditions for the vectors ${\vec k}$
and ${\vec l}$ imply that ${\vec l} = {\vec k} / (1+z)$, and hence the
affine parameters $s$ and ${\bar \lambda}$ are related by $s = {\bar
  \lambda}/(1+z)$.  The amplitude $A$ satisfies the differential equation
\be
d (\ln A) /d {\bar \lambda} = - {\bar \theta}/2,
\label{eq:de}
\ee
where ${\bar \theta} = \nabla_\alpha l^\alpha$ is the
expansion.  We choose the normalization of $A$ so that $A \approx
1/{\bar \lambda}$ for ${\bar \lambda} \to 0$ near ${\cal Q}$.

The energy flux at ${\cal P}$ can now be computed as $dE/(dt d^2A) =
T_{\alpha\beta} u^\alpha u^\beta =
A^2 (k_\alpha u^\alpha)^2 = A^2 (1+z)^{-2}$.  The luminosity at
${\cal Q}$ can be evaluated by integrating the energy flux over a
small sphere about $Q$ of radius ${\bar \lambda}$; this gives
$dE/dt = A^2 (l_{\alpha'} u^{\alpha'})^2 (4 \pi {\bar \lambda}^2) = 4
\pi$.  Combining these results yields ${\cal L} = (1+z)/A$.
Using the relation $s = {\bar \lambda}/(1+z)$ we can rewrite this as
\be
{\cal L} = (1 + z)^2 s \Delta(x,x')^{-1/2},
\label{eq:lumd}
\ee
where we have defined $\Delta  = A^2 {\bar \lambda}^2$.  This
quantity satisfies $\Delta \to 1$ as ${\cal Q} \to {\cal P}$ and also
from Eq.\ (\ref{eq:de}) satisfies the differential equation $d \ln
\Delta / (d {\bar \lambda}) = 2/{\bar \lambda} - {\bar \theta}$.
By comparing with Eq.\ (32) of Ref.\ \cite{Visser} we see that
$\Delta(x,x')$ is the van Vleck determinant\cite{DB60,Visser}.
Using the expansion $\Delta(x,x') = 1 + O(s^2)$ given in
Ref. \cite{Visser} and combining Eqs.\ (\ref{eq:redshift}) and
(\ref{eq:lumd}) gives a relation between redshift $z$ and luminosity
distance ${\cal L}$ of the form (\ref{eq:expansion0}),
where the coefficients are
\begin{eqnarray}
\label{eq:Aresult}
A(\theta,\varphi) &=& {1 \over (\nabla^\alpha u^\beta) k_\alpha
  k_\beta},  \\
\mbox{}
B(\theta,\varphi) &=& {2 \over (\nabla^\alpha u^\beta) k_\alpha k_\beta}
+ { (\nabla^\alpha \nabla^\beta u^\gamma) k_\alpha k_\beta k_\gamma
  \over 2 [ (\nabla^\alpha u^\beta) k_\alpha k_\beta ]^3}.
\label{eq:Bresult}
\end{eqnarray}

We now evaluate the averages over angles.  Using the definition $H_0 =
\langle A^{-1} \rangle$ together with $\langle k_\alpha k_\beta
\rangle  = (g_{\alpha\beta} + 4
u_\alpha u_\beta)/3$ yields the result (\ref{eq:H0ans}), using
$(\nabla^\alpha u^\beta) u_\alpha u_\beta = a^\alpha u_\alpha =0$.
Note that if we use the alternative angle-averaging definition $H_0^{-1} =
\langle A \rangle$ we instead obtain $H_0^{-1} = \langle (\theta/3 +
\sigma_{ij} n_i n_j + a_i n_i)^{-1} \rangle$, where $k^\alpha =
(1,n^i)$ in the local comoving frame at ${\cal P}$.
Evaluating this average treating the shear and acceleration as small
compared to the expansion yields
$
H_0 = \theta/3 - 2 \sigma_{ij} \sigma_{ij}/
  (5 \theta) - a_i a_i/\theta +
O(\sigma^4/\theta^3) + O[(a_i a_i)^2/\theta^3].
$
Thus the different averaging prescriptions give different answers.
However the fractional differences are of order $\sigma^2/\theta^2$,
which we have argued above is of order $\varepsilon$ and is small.

We now evaluate the angular average of the quantity
$A^{-3} B$.  We write this as $\langle A^{-3} B \rangle = I/2 + J$,
where
$
I \equiv \langle (\nabla^\alpha \nabla^\beta u^\gamma) k_\alpha k_\beta
k_\gamma \rangle
$
and
$J \equiv 2 \langle (\nabla^\alpha u^\beta k_\alpha k_\beta)^2
\rangle$.  For $J$ we obtain $J = 2 \langle ( \theta/3 + \sigma_{ij}
n_i n_j + a_i n_j)^2 \rangle = 2 \theta^2/9 + 4 \sigma_{ij}
\sigma_{ij}/15 + 2 a_i a_i/3$.
Using the formula $\langle k_\alpha k_\beta k_\gamma \rangle =
(g_{\alpha\beta} u_\gamma + g_{\alpha\gamma} u_\beta + g_{\beta\gamma}
u\alpha)/3 + 2 u_\alpha u_\beta u_\gamma$, we obtain for $I$ the formula
\begin{eqnarray}
3 I &=& u_\gamma \nabla^\gamma \nabla_\alpha u^\alpha
+ u_\gamma \nabla_\alpha \nabla^\gamma u^\alpha
+ u_\alpha \nabla_\gamma \nabla^\gamma u^\alpha
\nonumber \\
\mbox{} &&+ 6 u_\alpha u_\beta u_\gamma \nabla^\alpha \nabla^\beta u^\gamma.
\label{eq:3I}
\end{eqnarray}
We can rewrite the first term by commuting the covariant derivatives,
which gives $u_\gamma \nabla_\alpha \nabla^\gamma u^\alpha -
R_{\alpha\beta} u^\alpha u^\beta = \nabla_\alpha a^\alpha -
(\nabla_\alpha u_\gamma) \nabla^\gamma u^\alpha -
R_{\alpha\beta} u^\alpha u^\beta$.  Using the decomposition (\ref{eq:decompos})
this term can be written as
$
\nabla_\gamma a^\gamma - \theta^2/3 - \sigma_{\alpha\beta}
\sigma^{\alpha\beta} + \omega_{\alpha\beta} \omega^{\alpha\beta} -
R_{\alpha\beta} u^\alpha u^\beta.
$
Similarly the second term in Eq.\ (\ref{eq:3I}) evaluates to
$
\nabla_\gamma a^\gamma -\theta^2/3 - \sigma_{\alpha\beta}
\sigma^{\alpha\beta} + \omega_{\alpha\beta} \omega^{\alpha\beta}.
$
By differentiating twice the identity $u_\alpha u^\alpha=-1$, the
third term can be written as $- (\nabla_\alpha u_\beta) \nabla^\alpha
u^\beta$, which using the decomposition (\ref{eq:decompos}) evaluates
to
$
- \theta^2/3 - \sigma_{\alpha\beta} \sigma^{\alpha\beta} -
\omega_{\alpha\beta} \omega^{\alpha\beta} + a_\alpha a^\alpha.
$
Finally a similar manipulation of the fourth term shows that it
reduces to $- 6 a_\alpha a^\alpha$.  Combining these results
and using the Einstein equation to replace $R_{\alpha\beta} u^\alpha
u^\beta$ with $4 \pi (\rho + 3 p)$ gives
\begin{eqnarray}
I &=& {2 \over 3} \nabla_\alpha a^\alpha - {1 \over 3} \theta^2 -
\sigma_{\alpha\beta} \sigma^{\alpha\beta}
+ {1 \over 3} \omega_{\alpha\beta} \omega^{\alpha\beta}
\nonumber \\ \mbox{} &&
- {5 \over 3} a_\alpha a^\alpha - {4 \pi \over 3}
(\rho + 3 p).
\label{eq:Iresult}
\end{eqnarray}
Now combining the results for $I$ and $J$ and substituting into the
formula (\ref{eq:q0def}) for $q_0$ gives the result
(\ref{eq:q0result}).

Finally we note that using the alternative angular averaging
definition $q_0 = 1 - 2 H_0 \langle B \rangle$ would not change our
conclusions.  This average can be evaluated by
expanding the denominators in Eq.\ (\ref{eq:Bresult}) treating the
last three terms in Eq.\ (\ref{eq:decompos}) as small compared to the
expansion term, using the identity $\langle k_\alpha k_\beta
k_\gamma k_\delta k_\varepsilon \rangle =
16 u_\alpha u_\beta u_\gamma u_\delta u_\varepsilon/3
+ 16 g_{(\alpha\beta} u_\gamma u_\delta u_{\varepsilon)}/3
+ g_{(\alpha\beta} g_{\gamma\delta} u_{\varepsilon)}$, and performing
manipulations similar to those used above.  The
modifications to Eq.\ (\ref{eq:ans}) that result are: (i) changes to
the numerical coefficients of the shear squared and vorticity squared terms;
(ii) the addition of a term proportional to $H_0^{-3} u^\alpha
\nabla_\alpha (\sigma_{\beta\gamma} \sigma^{\beta\gamma})$ which is of
the same order as the shear squared term; and (iii) the addition of terms
that are smaller than the terms retained by one or more powers of the
small parameters
$\sqrt{\sigma_{\alpha\beta} \sigma^{\alpha\beta}}/\theta$ or
$\sqrt{\omega_{\alpha\beta} \omega^{\alpha\beta}}/\theta$.

This research was supported in part by NSF grant PHY-0140209.  I thank
David Chernoff, Saul Teukolsky, Henry Tye and Maxim Perelstein for
helpful conversations.

\newcommand{\apjl}{Astrophys. J. Lett.}
\newcommand{\aap}{Astron. and Astrophys.}
\newcommand{\cmp}{Commun. Math. Phys.}
\newcommand{\grg}{Gen. Rel. Grav.}
\newcommand{\lr}{Living Reviews in Relativity}
\newcommand{\mnras}{Mon. Not. Roy. Astr. Soc.}
\newcommand{\pr}{Phys. Rev.}
\newcommand{\prsl}{Proc. R. Soc. Lond. A}
\newcommand{\ptrsl}{Phil. Trans. Roy. Soc. London}

\end{document}